\documentclass{aipproc}

\usepackage{graphicx}

\layoutstyle{6x9}

\title[WKB States of Relativistic Extended Objects]%
{Minisuperspace, WKB, Quantum States of General Relativistic Extended Objects}

\author{S. Ansoldi}{
    address = {Dipartimento di Matematica e Informatica, Universit\`{a} di Udine, and
                I.N.F.N. Sezione di Trieste\\
                via delle Scienze, 206 - I-33100 Udine (UD) - ITALY
                },
                ,email = {ansoldi@trieste.infn.it},
                ,homepage = {http://www-dft.ts.infn.it/\~{}ansoldi}
}



\begin{abstract}
The dynamics of relativistic thin shells is a recurrent topic in the literature about the
classical theory of gravitating systems and the still ongoing attempts to obtain a coherent
description of their quantum behavior. Certainly, a good reason to make this system a
preferred one for many models is the clear, synthetic description given by Israel junction
conditions. Using some results from an effective Lagrangian approach to the dynamics of
spherically symmetric shells, we show a general way to obtain WKB states for the system;
a simple example is also analyzed.
\end{abstract}

\begin{document}

\maketitle

The study of the dynamics of an (infinitesimally)\footnote{Far from being only
idealizations of more realistic situations, shells have been
extensively used to build concrete models of many astrophysical and cosmological scenarios
(for a detailed bibliography, please, see the references in \cite{bib:ClQuG2002..19..6321A}).}
thin surface layer separating two
domains of spacetime is an interesting problem in General Relativity. The system can be
described in a very concise and geometrically flavored way using Israel's junction conditions
\cite{bib:NuCim1966.B44.....1I,bib:NuCim1967.B48...463I,bib:PhReD1991..43..1129I}.
Starting from these toeholds many authors have then tackled the problem of finding some hints
about the properties of the \textit{still undiscovered} quantum theory of
gravitational phenomena using shells as convenient models. In this context, just as
examples of what can be found in the literature, we quote the seminal works of Berezin
\cite{bib:PhLeB1990.241...402B} and Visser \cite{bib:PhReD1991..43...402V}, that date back
to the early nineties, or the more recent \cite{bib:JMPhA2002..17...979B,bib:PhReD2002..65064006R}
and references therein.

What we are going to shortly discuss in the present contribution is set in this last perspective
and suggests a semiclassical approach to define WKB quantum states for spherically symmetric shells.
This method has already been used in \cite{bib:ClQuG2002..19..6321A}.


Let us then consider a spherical shell (we refer the reader
to \cite{bib:PhReD1991..43..1129I} for very concise/clear background material and for definitions).
For our purpose the relevant result is equation (4) in \cite{bib:PhReD1991..43..1129I},
i.e. the junction conditions\footnote{Conventions are as in \cite{bib:Freem1970...1..1279W} and the definition
of the (quantities relevant to the concept of) extrinsic curvature can be found in
\cite{bib:PhReD1991..43..1129I,bib:Freem1970...1..1279W}.}
$
    K ^{-} _{ij} - K ^{+} _{ij} \propto S _{ij} - g _{ij} S / 2
    .
$
$K _{ij}$ is the extrinsic curvature of the shell and can have different values on the two sides
($+$ and $-$ spacetime regions) of it. $S _{ij}$ is the stress energy tensor describing the
energy/matter content of the shell ($S$ is its trace). For a spherical shell these equations
reduce to the single condition
\begin{equation}
    \epsilon _{-} ( \dot{R} ^{2} + f _{-} (R) ) ^{1/2}
    -
    \epsilon _{+} ( \dot{R} ^{2} + f _{+} (R) ) ^{1/2}
    =
    M (R) / R
    ,
\label{eq:sphjuncon}
\end{equation}
where $f _{\pm} (r)$ are the metric functions in the static coordinate systems adapted to the
spherical symmetry of the $4$-dimensional spacetime regions joined
across the shell; $\epsilon _{\pm} =  +1, -1$ are signs and $R$ and $M (R)$ are the the radius
(a function of the proper time $\tau$ of a co-moving observer\footnote{As usual an
overdot, ``$\dot{\quad}$'',denotes a total derivative with respect to $\tau$.}) and the
matter content (what remains of $S _{ij}$) of the spherical shell, respectively.
As shown for example in
\cite{bib:NuPhB1990.339...417G,bib:ClQuG1997..14..2727S,bib:Thes.1994....TriestA}
the above equation can be obtained from an effective Lagrangian\footnote{For
a more general and deeper discussion see
\cite{bib:PhReD2000..62044025K} and references therein or also, in addition,
\cite{bib:JMaPh2001..42..2590G,bib:Thes.2004....TriestS}.} $L _{\mathrm{EFF}} (R , \dot{R})$,
as a first integral of the second order Euler-Lagrange equation. From $L _{\mathrm{EFF}} (R , \dot{R})$
the momentum conjugated to the single degree of freedom $R$ can be obtained as usual,
$P (R, \dot{R}) = \partial L _{\mathrm{EFF}} (R , \dot{R}) / (\partial \dot{R})$.
We are not interested in the explicit form of $P$ here, we just point out that it is
a function of $R$ and $\dot{R}$ highly \emph{non-linear} in $\dot{R}$. This non-linearity
spoils the natural and simple interpretation of the momentum than we know from classical
analytical mechanics. Nevertheless we can still solve \eqref{eq:sphjuncon} for the classically
allowed trajectories of the shell, using a \emph{standard} analogy with the motion of a
\textit{unitary mass particle with zero energy in an effective potential}
\cite{bib:PhReD1987..36..2919T,bib:PhReD1987..35..2961G,bib:PhReD1989..40..2511S}.
This gives $\dot{R}$ as a function of $R$ and, substituting this expression for
$\dot{R}$ in $P (R, \dot{R})$, we obtain the \textit{conjugated momentum on
a solution of the classical equations of motion}. In what follows we are going to indicate
the momentum \emph{evaluated on a classical trajectory} as $P (R ; {\mathcal{S}})$: this
emphasizes that it is a function of $R$, that it depends on the set ${\mathcal{S}}$ of the other
parameters of the problem, but, of course, it is not a function of $\dot{R}$.

\begin{figure}
    \includegraphics[width=8cm]{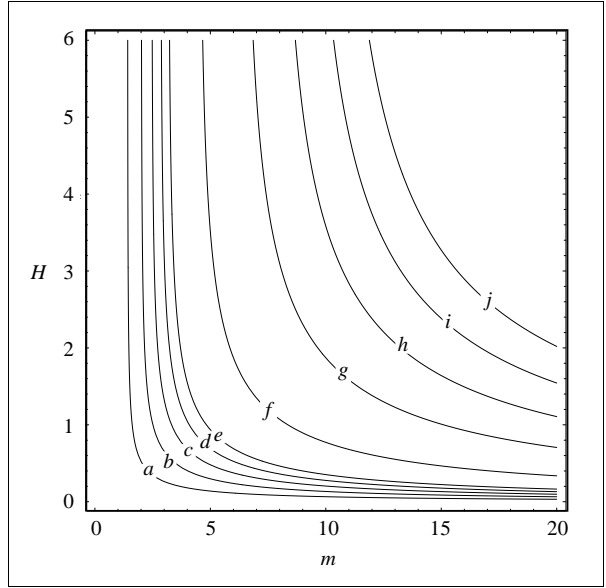}
\caption{\label{fig:actlev}{\small{}Plot of some WKB levels for the example discussed in the text.
The curves $a$, $b$, $c$, $d$, $e$, $f$, $g$, $h$, $i$, $j$ correspond to the quantum numbers $1,2,3,4,5,10,20,30,40,50$, respectively.}}
\end{figure}

By integrating  the expression for $P (R ; {\mathcal{S}})$ on a classically allowed
trajectory with turning points $R _{\mathrm{MIN}}$ and $R _{\mathrm{MAX}}$, we can compute the
value of the classical action for that trajectory. It is a function of the set of parameters
characterizing the matter content and of the geometry, ${\mathcal{S}}$, and WKB quantum states of the
system can now be defined as states for which the above action is a multiple of the quantum
\begin{equation}
    S ( {\mathcal{S}} )
    =
    2 \int _{R _{\mathrm{MIN}}} ^{R _{\mathrm{MAX}}}
        P (R ; {\mathcal{S}}) d R
    \sim n,
    \quad n = 0 , 1 , 2 , 3 , \dots{}.
\label{eq:BohSom}
\end{equation}


In quantum gravity we expect to have a theory that selects some geometries from the set
of all possible ones consistently with quantum dynamics. In our discussion we have limited the
treatment to a \emph{minisuperspace approximation}, but, indeed, we see that the quantization condition
\eqref{eq:BohSom} does select only some of the possible geometries, those in which the parameters of
the model are related by the quantum number $n$ as in \eqref{eq:BohSom}.
We can see this more explicitly in the following simple model: a dust shell ($M (R) = m$)
separating two domains of anti-de Sitter spacetime with the same cosmological constant
($f _{-} (r) = f _{+} (r) = f (r) = 1 + r ^{2} / H ^{2}$):
it is possible to prove that a non-trivial junction
of the two spacetimes can be performed, although we are not going to discuss this aspect here
nor we are going to describe the resulting global spacetime structure. In this case the set of
parameters is ${\mathcal{S}} = \{ m , H \}$ and the quantization condition \eqref{eq:BohSom} becomes
$S ( m , H ) \sim n$, $n = 0 , 1 , 2 , 3 , \dots{}$. Some levels are plotted in figure
\ref{fig:actlev} and clearly show that given one of the parameters (say $m$) only a discrete set
of values for the other $H$ is allowed: thus the quantization condition restricts the possible
values that can be given to the parameters that characterize the spacetime geometry and/or
the matter content of the shell. This is consistent with the general picture of a quantum
gravitational scenario.

Other applications of this general approach are presently under study. In particular,
generalizations to higher dimensions \cite{bib:none.when........AGI} could be relevant,
for example, in the context of brane cosmological models.


\begin{thebibliography}{19}
\expandafter\ifx\csname natexlab\endcsname\relax\def\natexlab#1{#1}\fi
\providecommand{\enquote}[1]{``#1''}
\expandafter\ifx\csname url\endcsname\relax
  \def\url#1{\texttt{#1}}\fi
\expandafter\ifx\csname urlprefix\endcsname\relax\def\urlprefix{URL }\fi

\bibitem[{Israel}(1966)]{bib:NuCim1966.B44.....1I}
{Israel}, W., \emph{Nuovo Cimento B}, \textbf{44}, 1 (1966).

\bibitem[{Israel}(1967)]{bib:NuCim1967.B48...463I}
{Israel}, W., \emph{Nuovo Cimento B}, \textbf{48}, 463 (1967).

\bibitem[{Barrabes} and {Israel}(1991)]{bib:PhReD1991..43..1129I}
{Barrabes}, C., and {Israel}, W., \emph{Phys. Rev. D}, \textbf{43}, 1129--1142
  (1991).

\bibitem[{Ansoldi}(2002)]{bib:ClQuG2002..19..6321A}
{Ansoldi}, S., \emph{Class. Quantum Grav.}, \textbf{19}, 6321--6344 (2002).

\bibitem[{Berezin}(1990)]{bib:PhLeB1990.241...402B}
{Berezin}, V.~A., \emph{Phys. Lett. B}, \textbf{241}, 194--200 (1990).

\bibitem[{Visser}(1991)]{bib:PhReD1991..43...402V}
{Visser}, M., \emph{Phys. Rev. D}, \textbf{43}, 402--409 (1991).

\bibitem[{Berezin}(2002)]{bib:JMPhA2002..17...979B}
{Berezin}, V., \emph{J. Mod. Phys. A}, \textbf{17}, 979--988 (2002).

\bibitem[{Corichi} et~al.(2002)]{bib:PhReD2002..65064006R}
{Corichi}, A., {Cruz-Pacheco}, G., {Minzoni}, A., {Padilla}, P., {Rosenbaum},
  M., {Ryan Jr.}, M.~P., {Smyth}, N.~F., and {Vukasinac}, T., \emph{Phys. Rev.
  D}, \textbf{65}, 064006(1--13) (2002).

\bibitem[{Misner} et~al.(1970)]{bib:Freem1970...1..1279W}
{Misner}, C.~W., {Thorne}, K.~S., and {Wheeler}, J.~A., \emph{"Gravitation"},
  W. H. Freeman and Company, 1970.

\bibitem[{Farhi} et~al.(1990)]{bib:NuPhB1990.339...417G}
{Farhi}, E., {Guth}, A.~H., and {Guven}, J., \emph{Nucl. Phys. B},
  \textbf{339}, 417--490 (1990).

\bibitem[{Ansoldi} et~al.(1997)]{bib:ClQuG1997..14..2727S}
{Ansoldi}, S., {Aurilia}, A., {Balbinot}, R., and {Spallucci}, E., \emph{Class.
  Quantum Grav.}, \textbf{14}, 2727--2755 (1997).

\bibitem[{Ansoldi}(1994)]{bib:Thes.1994....TriestA}
{Ansoldi}, S., \emph{Graduation Thesis (in Italian)}, University of
  Trieste (1994).

\bibitem[{Hajicek} and {Kijowski}(2000)]{bib:PhReD2000..62044025K}
{Hajicek}, P., and {Kijowski}, J., \emph{Phys. Rev. D}, \textbf{62},
  044025(1--5) (2000).

\bibitem[{Gladush}(2001)]{bib:JMaPh2001..42..2590G}
{Gladush}, V.~D., \emph{J. Math. Phys.}, \textbf{42}, 2590--2610 (2001).

\bibitem[{Sindoni}(2004)]{bib:Thes.2004....TriestS}
{Sindoni}, L., \emph{Graduation Thesis (in Italian)}, University of
  Trieste (2004).

\bibitem[{Berezin} et~al.(1987)]{bib:PhReD1987..36..2919T}
{Berezin}, V.~A., {Kuzmin}, V.~A., and {Tkachev}, I.~I., \emph{Phys. Rev. D},
  \textbf{36}, 2919--2944 (1987).

\bibitem[{Blau} et~al.(1987)]{bib:PhReD1987..35..2961G}
{Blau}, S.~K., {Guendelman}, E.~I., and {Guth}, A.~H., \emph{Phys. Rev. D},
  \textbf{35}, 1747--1766 (1987).

\bibitem[{Aurilia} et~al.(1989)]{bib:PhReD1989..40..2511S}
{Aurilia}, A., {Palmer}, M., and {Spallucci}, E., \emph{Phys. Rev. D},
  \textbf{40}, 2511--2518 (1989).

\bibitem[{Ansoldi} et~al.({In preparation})]{bib:none.when........AGI}
{Ansoldi}, S., {Guendelman}, E.~I., and {Ishihara}, H. ({In preparation}).

\end{thebibliography}

\end{document}